\documentclass[reprint,groupedaddress,superscriptaddress,amsmath,amssymb,aps,twocolumn,floatfix, prx,table]{revtex4-2}
\usepackage{graphicx}
\usepackage{rotating}
\usepackage{dcolumn}
\usepackage{tikz}
\usepackage{appendix}
\definecolor{mygreen}{rgb}{0.2, 0.8, 0.2}
\usepackage[colorlinks=true, linkcolor=blue, citecolor=mygreen, urlcolor=red]{hyperref}
\usepackage{hhline}
\usepackage{listings} 
\usepackage{booktabs} 
\usepackage{bm}
\usepackage{multirow}
\usepackage{array}
\usepackage{comment}
\usepackage{colortbl}
\usepackage{comment}
\usepackage[compat=1.1.0]{tikz-feynman}
\usepackage{braket}
\usepackage{simpler-wick}
\usepackage{MnSymbol}
\usepackage{eucal}
\usepackage{epstopdf}
\usepackage{float}
\usepackage{svg}
\usepackage{dcolumn}
\usepackage{eqnarray, amsmath}
\usepackage{amsthm}
\usepackage{scalerel}
\usepackage{stackengine,wasysym}
\usepackage{yfonts}
\usepackage{bbm}
\usepackage{mathbbol}
\usepackage{amsfonts}
\usepackage{dsfont}
\usepackage{titlesec}
\usepackage{xr-hyper}
\usepackage{tikz}   

\usepackage{amsmath}
\usepackage{etoolbox}  

\makeatletter

\counterwithout*{figure}{section}
\@removefromreset{figure}{section}

\counterwithout*{table}{section}
\@removefromreset{table}{section}
\makeatother

\newenvironment{singlecolfigure}{%
  \onecolumngrid
  \begin{figure*}
}{%
  \end{figure*}
  \twocolumngrid
}
\newenvironment{singlecoltable}{%
  \onecolumngrid
  \begin{table*}
}{%
  \end{table*}
  \twocolumngrid
}
\newcommand{\twocolcaption}[1]{%
  \par\vspace{0.5em}%
  \noindent
  \begin{minipage}[t]{0.94\textwidth}  
    \raggedright
    \small #1
  \end{minipage}
}

\begin{document}
\definecolor{lightblue}{RGB}{173, 250, 230}
\title{Semi-Deterministic and Stochastic Sampling of Feynman Diagrams with $1/\text{N}_{f}$ Expansions}
\author{Boyuan Shi}
\email{boyuanshi0607@gmail.com}
\affiliation{Blackett Laboratory, Imperial College London,
London SW7 2AZ, United Kingdom}

\begin{abstract}
We introduce a family of (semi) bold-line series, assisted with $1/\text{N}_{f}$ expansions, with $\text{N}_{f}$ being the number of fermion flavours. If there is no additional $\mathrm{N}_{f}$ cut, the series reduces to the RPA series in the density-density channel, complementary to the particle-hole and particle-particle channels introduced in [Phys. Rev. B \textbf{102}, 195122 (2020)]. To address the very localized integrands in diagrammatic Monte Carlo, we introduced an innovative VEGAS-RG-MCMC sampling method, where we found an significant decrease of autocorrelation time without the usage of state-of-art many-configuration MCMC (MCMCMC) method while the combination of both is also straightforward. We performed extensive benchmarks for density, energy and pressure with $\mathrm{t}-\mathrm{t}'$ $\mathrm{SU}(\mathrm{N}_{f})$ Hubbard model on square lattice and honeycomb lattices over a wide range of numerical methods. For benchmark purposes, we also implement bare-$\mathrm{U}$ symmetry-broken perturbation series for the 2D SU(2) Hubbard model on the honeycomb lattice, where we found encouraging results from weak to intermediate couplings.
\end{abstract}
\maketitle
\section{INTRODUCTION}
Accurately simulating quantum many-body systems to understand and predict materials properties has high academic and industrial values. The major bottleneck for real material simulations is the exponential wall problem for large systems. Finite-size effects are critical in many aspects of material simulations, as they can significantly influence the accuracy of results. In particular, small scale simulations may not fully capture long-range correlations. At the vicinity of phase transitions, divergence of correlation length makes it difficult to extrapolate to the thermodynamic limit. The other issue is the inclusion of multiple orbitals and interaction matrix elements. Properly handling them is already challenging for methods that only require limited computational resources, such as GW and its extensions \cite{GW_Aryasetiawan_1998}.

Field theoretical approaches among concurrent methods \cite{DMFT_review, DMRG_review, tensor_networks_review, fRG_review, QMC_review, Hirsch_DQMC, WSSLGS_DQMC, NLCE, GW_Aryasetiawan_1998, DFT_review} usually could be applied to sufficiently large systems. Unless at extremely low temperatures or ground state for some special cases, real materials simulations are typically free from both the UV and IR divergence, and perturbation series based on interaction strength could acquire a finite radius of convergence. The very primordial idea to push it to higher orders stages a comeback with the advent of computing power, with numerous studies showing its unprecedented accuracy over many models that are close to real materials.

4With the finite-size effects seemingly resolved by this class of methods, the simple change of model from one-band to two-bands considerably circumscribes the applicability due to the additional exponential growth of computational costs. In particular, although the family of diagrammatic Monte Carlo (DiagMC) methods have superior accuracy over other field theoretical methods, it suffers from such problems.

For methods within the DiagMC family, computational costs for complicated lattice geometries and more interaction matrix elements increase exponentially with the state-of-arts principal minor algorithms. The very recent GPU speed-up based on adaptions of a combinatorial algorithm for determinants \cite{CoS_GPU_speed_up} provides a promising routine but algorithmic adaptions for different lattice and interaction configurations are highly nontrivial. 

The other major problem of diagrammatic Monte Carlo is the enlargement of its radius of convergence and the reliability of re-summation. The region of analyticity of any power series is delineated by a circle. For example, even if there is no phase transition in weakly-interacting fermi liquids, the negative interaction value on the other side might well correspond to a superfluid phase transition or BCS instability. Therefore, re-summation must be invoked at almost all the times. That does not appear to constitute a significant impediment given that sufficiently high orders of the perturbation series can be reached. Indeed, this has been extensively examined in 2D and 3D Hubbard models on square and cubic lattices. However, two underlying assumptions must be invoked. That the maximum truncation order around nine to ten is generally sufficient and the analytical structure remains sufficiently accommodating for the functionality to persist when generalizing to real materials. The breakdown of these assertions may plausibly manifest in previously studied models, while series with intrinsically larger radius of convergence are useful in uncovering strongly correlated physics.

Given that (semi)-bold-line series has better convergence properties, we return to RPA series and proceed with an attempt for $1/\mathrm{N}_{f}$ expansions, with $\mathrm{N}_{f}$ being the number of fermion flavours. The initial impetus for this idea stemmed from the observation that $\mathrm{N}_{f}$ is independent of the dynamical parameters of the system and can be large even for real physical systems.

For real materials that $1/\mathrm{N}_{f}$ expansions would apply, there is a family of materials that Hund and exchange interactions are weak, typically in compounds when the $d$, $f$ orbitals are filled, \textit{e.g.} $\mathrm{Zn}\mathrm{O}$. Such situations are common in wide band-gap semiconductors, while commonly adapted approaches are still based on DFT \cite{DFT_review} or GW \cite{GW_Aryasetiawan_1998}.

Compared to other bold-line schemes, \textit{e.g.} $\text{G}^{2}\text{W}$ \cite{G2W}, $\text{G}^{2}\Gamma$ \cite{G2Gamma}, \textit{etc}, the scheme we introduced does not require self-consistency at the DiagMC level but only when preparing tabulated bold-line building blocks. It is commonly agreed in the community that there is no misleading convergence for semi-bold diagrammatic series \cite{semi_bold_justification}. Compared to order-by-order RPA series, new features of the $1/\mathrm{N}_{f}$ expansions are that they change the diagram topologies by assigning higher priorities to diagrams with more closed fermion loops. With the $\mathrm{N}_{f}$ cut removed, the series returns to the order-by-order RPA series in the density-density channel while all the other bold-line series in the literature are in the particle-particle and particle-hole channels.

Besides a purely academic exposition for a new series, we applied the method to study Hubbard-type models beyond the square lattice. It is intriguing to ask if better convergence on honeycomb lattice may be achieved by adapting existing bold-line algorithms, in particular RPADet  \cite{RPADet} or CoS \cite{CoS}, the only two that admit exponential algorithms. For RPADet, it is not easy to see how symmetry broken terms akin to those in \cite{symmetry_broken_perturbation_series} and additional fermion flavors could be added in the way that are consistent with the fast algorithm in \cite{RPADet}. For CoS, it is indeed mentioned in \cite{CoS} that there can be a CoS-GW algorithm that might solve the fermion flavours problem. However, no applications to any physical systems have been reported yet, especially for those that exhibit symmetry breaking of any kind.

Besides the obvious perspective of providing cross benchmarks, we aim to use this new series to study strongly-correlated physics. This would include both conventional $\mathrm{SU}(2)$ models and more exotic $\mathrm{SU}(\mathrm{N}_{f}>2)$ models. For $\mathrm{SU}(2)$ models, a multitude of outstanding questions remains to be addressed in systems with long-range interactions, frustrated lattice geometries, and stacked multi-layers structures. Such systems engender nontrivial complications for both numerically exact methods such as determinant quantum Monte Carlo and density-matrix renormalization group or heuristic methods such as dynamical mean filed theories and its extensions. $1/\mathrm{N}_{f}$ expansions are anticipated to mitigate the limitations of these approaches in such contexts.

For $\mathrm{SU}(\mathrm{N}_{f})$ models, extensive numerical and experimental studies are still largely blank. Those models can directly simulate a box of atoms or molecules with high nuclear spins under external potentials, where there have been several preliminary experimental studies on ultracold alkali atoms \cite{SU_N_Cold_atom_antiferromagnetic_transition, EoS_SU_N_Cold_atom} in optical lattices. In periodic solids, such situations occur naturally in monolayer and stacked-bilayer graphenes, transition metals oxides, \textit{etc}. Given that almost none of the real materials that have multiple orbits is on a square lattice with the ideal on-site Hubbard interactions, it is still a highly non-trivial task for the state-of-art CoS algorithm \cite{CoS} to generalize to those scenarios.

Despite in this article we only present results with the density-density saddle-point, switching to mixed channels would share identical set of diagrams and employs analogous combinatorial algorithms. The more intriguing case is of course the cross of quantum phase transitions that are not adiabatically connected to this type of saddle point. We have seen evidence for the $1/\mathrm{N}_{f}$ expansion to cover those cases while the optimal truncation order can be learned and generalized. 

As our second purpose, we also wish to reduce computational costs for more complicated situations. Even the leading-order corrections in $1/\mathrm{N}_{f}$ could furnish stable refinements to random-phase approximations, while the required CPU time is within minutes. Although the sign problem in equilibrium is manageable by other DiagMC algorithms, they would all struggle in non-equilibrium due to the additional exponential complexity in reaching long times. In disordered systems, many-body localizations would require an ensemble of thousands of systems copies generated from the distribution of disorders, where other DiagMC methods would require enormous computational resources.

The article proceeds as follows. In Sec. \ref{sec:MCMC}, we give an overview of the Markov-Chain Monte Carlo methods we use. In Sec. \ref{sec:G_Sigma}, we derive the $1/\mathrm{N}_{f}$ expansions in the framework of $G-\Sigma$ functional theory. In Sec. \ref{sec:leading_orders}, we give leading-order computations, higher-orders diagram generations, and low-rank decompositions for diagrams building blocks. In Sec. \ref{sec:examples}, we apply the method to study $\mathrm{t}-\mathrm{t}'$ $\mathrm{SU}(2)$ and $\mathrm{SU}(6)$ Hubbard model on square and honeycomb lattices.

\section{MARKOV-CHAIN MONTE CARLO METHODOLOGIES} \label{sec:MCMC}
There are in general two classes of algorithms in performing DiagMC, where the first treats the space of topologies and space(momentum)-time continuum vertices in the equal status and perform random walk in this space \cite{VANHOUCKE201095}. The second then firstly pre-generates all diagrams (or employs the determinant trick \cite{det_1,det_2,det_3,Rossi_DDMC,RPADet,CoS,TTCross_FeyDiag,fast_principle_minor_DDMC}) and merely performs random walks in the space of continuous parameters \cite{Chen_Haule, Rossi_DDMC, CoS}. We will discuss the first approach in this section, suitable for low to intermediate orders. Further optimizations will be discussed later.

To quantify the sign problem, we define a norm as follows. For given $\mathcal{N}$ diagrams, we insert $\mathcal{M}$ boards between them and sum diagrams deterministically within adjacent boards. We perform random walks in $(\mathcal{M}+1)$ group indices. Minimum of computational complexity then corresponds to the minimum of
\begin{equation}
    \mathcal{S}(\mathcal{Q},\mathcal{P})=\mathcal{C}(\mathcal{Q},\mathcal{P})\mathcal{A}(\mathcal{Q},\mathcal{P})^{2}\tau_{f}(\mathcal{Q},\mathcal{P}),
\end{equation}
where $\mathcal{A}(\mathcal{Q},\mathcal{P})$ is the integral of the absolute value of the function with partition $\mathcal{Q}$ recording boards positions and permutation $\mathcal{P}$ permuting diagrams ordering. $\mathcal{C}(\mathcal{Q},\mathcal{P})$ is the computational costs due to deterministic summation, which can be estimated during the execution. $\tau_{f}(\mathcal{Q},\mathcal{P})$ is the integrated auto-correlation time (IACT). In practice, the order of magnitude of the norm function and the IACT could be estimated with much less Monte Carlo iterations (\textit{e.g.} the blocking method \cite{blocking}), which renders minimization algorithms applicable, e.g. stimulated annealing \cite{stimulated_annealing}.

To evaluate high-dimensional integrals arisen from Feynman diagrams, we combine the VEGAS algorithm \cite{VEGAS_Original,Vegas_plus} and conventional Metropolis-Hastings algorithm for Markov-chain Monte Carlo \cite{robert2004monte, martino2018independent}. Consider
\begin{equation}
    I=\int_{\boldsymbol{x}\in\Omega} f(\boldsymbol{x})\,\mathrm{d}\boldsymbol{x}.
\end{equation}
The very peaky nature of the integrand could be retrieved to some extent by employing the VEGAS map. That is, a separable variable transformation, $\{y_{i}(x_{i})\}_{i\in\{1,...\mathrm{dim}\}}$ is introduced by learning the integrand via iterations of plain Monte Carlo evaluations. In the transformed variable space,
\begin{equation}
    I=\int_{\boldsymbol{y}\in[0, 1]^{\otimes\mathrm{dim}}} f(\boldsymbol{x}(\boldsymbol{y}))J(\boldsymbol{y})\,\mathrm{d}\boldsymbol{y}
\end{equation}
can be estimated by separating the sign and absolute value parts:
\begin{equation}
    I=\mathcal{N}\int_{\boldsymbol{y}\in[0, 1]^{\otimes\mathrm{dim}}} \mathrm{sign}\left[\tilde{f}_{\mathcal{N}}(\boldsymbol{y})\right]|\tilde{f}_{\mathcal{N}}(\boldsymbol{y})|\,\mathrm{d}\boldsymbol{y},
\end{equation}
where $\mathrm{dim}$ is the dimension, normalization factor $\mathcal{N}=\int_{\boldsymbol{y}\in[0, 1]^{\otimes\mathrm{dim}}}|f(\boldsymbol{x}(\boldsymbol{y}))J(\boldsymbol{y})|$, and $\tilde{f}_{\mathcal{N}}(\boldsymbol{y})=f(\boldsymbol{x}(\boldsymbol{y}))J(\boldsymbol{y})/\mathcal{N}$. The auto-correlation timescale for the average sign is usually short, since $\mathrm{sign}(f)$ is a very good Monte-Carlo measurement. The normalization constant can be estimated by a plain Monte-Carlo integration. However, in many cases, there are significant outliers in the samples, which significantly downgrades the reliability of samples. 

One way to address such issues is the usage of many-configuration Markov-Chain Monte Carlo (MCMCMC) \cite{MCMCMC}, which proposes multiple moves at each steps following graphic patterns. This approach is efficient when there is a parallel machine. This development is generic to any underlying distribution but does not wisely make use of the internal structures of the specific problem at hand. 

One can observe that there are multiple scales for any integrand involving Feynman diagrams, temperatures, system size, interaction strengths, filling, etc. The difficulty of integration decreases exponentially fast if vertices can be compactified. This motivates the design of MCMC based on renormalization-group transformations, where for the normalization constant, we design a chain or multiple chains of measurements to gradually simplify the problem. Let us denote $g(\boldsymbol{y})_{\beta,N,U,\mu,...}=|f(\boldsymbol{x}(\boldsymbol{y}))J(\boldsymbol{y})|_{\beta,N,U,\mu,...}$ after the VEGAS map. We have introduced subscripts to denote scales. For example, if the current integrand is at temperature $\beta$, we introduce an auxiliary function at higher-temperature $g'(\boldsymbol{y})_{\beta',N,U,\mu,...}$, where $\beta'<\beta$. We measure two observables, 
\begin{equation}
    O_1 = \left\langle\frac{g}{g+g'}\right\rangle_{g+g'},\quad O_2=\left\langle\frac{g'}{g+g'}\right\rangle_{g+g'}.
\end{equation}
And thus $\int g/\int g' = O_1/O_2$. Both $O_1$ and $O_2$ are estimated using the MCMC method, with the chain length much shorter than that in the sign estimate due to strict positivity. Now, we apply a further transformation with $g''$ at $\beta''<\beta'$ to estimate $g'$. $\int g'/\int g'' = O_3/O_4$. This chain ceases with the auxiliary function such that a direct VEGAS integration is sufficient to estimate its integral easily. Similar procedures can be applied against interactions, system size and chemical potentials as well. We show a practical example in Appendix. \ref{app:RG_MCMC}.

Convergence for paralleling multiple Markov chains was checked by the blocking method \cite{blocking} and the Gelman-Rubin statistics \cite{Gelman_Rubin_R}. We give more details of the parameters used for MCMC in Appendix. \ref{app:MCMC_para}.

For models with contact interactions, screened interactions contain a contact part with bare $\mathrm{U}$ and a dynamical part, which enlarges the Monte-Carlo variance. The switch between them introduces empty time vertices on which the integrand does not explicitly depend. One may observe that the compensation trick introduced in \cite{Compensation_Trick} just amounts to fill in those empty nodes with the suggested vertex function in \cite{Compensation_Trick}. We showed in the subsequent sections that one may further integrate out internal time vertices with the aid of low-rank decomposition.
\section{THEORETICAL FRAMEWORK} \label{sec:G_Sigma}
We consider the $\mathrm{SU}(\mathrm{N}_{f})$ Hubbard model in generic lattices. The (normal ordered) Hamiltonian is
\begin{equation}
\begin{aligned}
H=&-\sum_{\langle i, j\rangle, m}t_{ij}\,(\bm{c}^{\dagger}_{i,m}\bm{c}_{j, m}+\text{H.c.})-\mu\sum_{i,m}\bm{c}^{\dagger}_{i,m}\bm{c}_{i, m}\\
&+\frac{1}{2N_{f}}\sum_{i, j, m, m^{\prime}}v_{\alpha\beta}(i, j) c^{\dagger}_{i,m,\alpha}c^{\dagger}_{j,m^{\prime},\beta}c_{j,m^{\prime},\beta}c_{i,m,\alpha},
\end{aligned}
\end{equation}
where $i$, $j$ denote unit cell positions, $\alpha$, $\beta$ denote lattice sites in one unit cell and $m$, $m^{\prime}$ are flavour indices, and bold symbols for $c$ and $c^{\dagger}$ denote a vector of spin-up and spin-down creation and annihilation operators. Hopping integrals and interactions are not specified at this stage but could be of various types, \textit{e.g.} NN in the honeycomb lattice.

In the presence of spontaneous spin-rotation symmetry breaking, Green's functions are not homogeneous over different spins. As a scrupulous choice, we reformulate the fermionic path integral in the $\text{G}-\Sigma$ functional theory \cite{SYK_soft_modes,Sachdev_large_N_critical_FS}, where the field degrees of freedom is the spin-averaged bosonic fields. The action is given by
\begin{widetext}
\begin{equation}
\begin{aligned}
    S=&\sumint_{\tau,\tau',i,j} \bar{\bm{\psi}}_{m}(i, \tau)\delta(\tau-\tau')\left[(\partial_{\tau}-\mu)\delta_{ij}+h_{ij}\right]\bm{\psi}_{m}(j,\tau')+\frac{N_{f}}{2}\sumint_{\tau,\tau',i, j}\delta(\tau-\tau')G_{\alpha\alpha}(i, i, \tau,\tau^{\prime})G_{\beta\beta}(j, j, \tau,\tau^{\prime})v_{\alpha\beta}(i, j)\\
    &-N_{f}\sumint_{\tau,\tau',i,j}\Sigma_{\beta\alpha}(j,i;\tau',\tau)\left[G_{\alpha\beta}(i, j;\tau,\tau')-\frac{1}{N_{f}}\sum_{m=1}^{N_{f}}\psi_{\alpha,m}(i\tau)\bar{\psi}_{\beta,m}(j,\tau')\right],
\end{aligned}
\end{equation}
\end{widetext}
where $h_{ij}$ are hopping matrix elements. We have introduced a pair of bilinear fields, G and $\Sigma$ to reformulate the original action \footnote{The convention is $Z=\int \mathcal{D}\Sigma\,\mathcal{D}G\,e^{-S}$, where $\mathrm{Z}$ is the grand-canoical partition function.}. We now introduce the compact index notation $1=(\tau,i,\alpha)$. After integrating out the fermionic fields, the effective action for $G$ and $\Sigma$ is
\begin{equation}
\begin{aligned}
\frac{S[G, \Sigma]}{N_{f}}=&-\mathrm{Tr}\ln[G_{0}^{-1}-\Sigma]-\sumint_{1,2}\Sigma(1,2)G(2,1)\\
&+\frac{1}{2}\sumint_{1,2,3,4}G(1, 2)G(3, 4)v(2, 3)\delta_{1,2}\delta_{3,4}.
\end{aligned}
\end{equation}
To the leading order in $1/\mathrm{N}_{f}$, the saddle point equation $\delta S/\delta G=0$ and $\delta S/\delta \Sigma=0$ is a self-consistent Hartree equation and the Dyson's equation, namely
\begin{equation}
     \Sigma_{\star}(1,2)=\sumint_{3,4}v(1,3)G(3,4)\delta_{1,2}\delta_{3,4}
\end{equation}
and 
\begin{equation}
    G_{*}=[G_{0}^{-1}-\Sigma_{\star}]^{-1}.
\end{equation}
This coincides with the naïve power counting, with the Feynman diagram being the Hartree diagram,
\begin{equation*}
\begin{tikzpicture}
\begin{feynman}
\vertex (b) at (14.5, 2.4);
\vertex (c) at (14.5, 1.1);
\vertex (d) at (14.5, 0);
\diagram*[scale=1]
{(b) -- [half right, line width=2] (c), 
(c) -- [half right, line width=2] (b), 
(c) -- [photon, color=blue] (d)};
\vertex [left = 0.15cm of d];
\vertex [right = 0.15cm of d];
\vertex [below = 0.2cm of d];
\end{feynman}
\end{tikzpicture}
\end{equation*}
We expand the $\text{G}-\Sigma$ action around the saddle point to quadratic order with the fluctuation field $G=G_{*}+\delta G$, $\Sigma=\Sigma_{*}+\delta\Sigma$ using the functional Taylor expansion, where the four second-order derivatives can be easily computed. Propagators between bi-linear fields can then be computed, where we denote them as $G_{\delta\Sigma\delta\Sigma}$ and $G_{\delta\Sigma\delta G}$. One may readily realize that $G_{\delta\Sigma\delta\Sigma}$ plays the same role as the screened interactions in the random-phase approximation (RPA)
$G_{\delta\Sigma\delta\Sigma}(1,2;3,4)\equiv W(1,3)\delta_{1,2}\delta_{3,4}$, where $\delta W(1,3)=W(1,3)+v(1,3)$ is graphically represented as
\begin{equation*}
\begin{tikzpicture}
\begin{feynman}
\vertex (b) at (14.5, 1.4);
\vertex (c) at (16, 1.4);
\vertex (d) at (17.5, 1.4);
\vertex (e) at (19, 1.4);
\diagram*[scale=1]
{(b) -- [photon, color=blue] (c), 
(c) -- [fermion, half right, line width=2] (d),
(d) -- [fermion, half right, line width=2] (c), 
(d) -- [photon, line width=2, color=red] (e)};
\vertex [left=0.3em of b] {\(1\)};
\vertex [right=0.3em of e] {\(3\)};
\end{feynman}
\end{tikzpicture}
.
\end{equation*}
The thin wiggle line is bare interaction vertex, $v$, bold wiggle line is $W$, and the bold solid directed line is $G_{*}$. Beyond the quadratic order, one may expand the effective action as:
\begin{equation}
S=N_{f}S_{\star}+\sum_{n=2}^{\infty}N_{f}S_{n}[\delta\Sigma, \delta G].
\end{equation}
After rescaling the fields and expanding the trace log, interacting parts in the action are given as
\begin{equation}
    S_{\text{I}}=\sum_{n=3}\frac{1}{n}N_{f}^{1-n/2}\mathrm{Tr}\left[(G_{\star}\delta\Sigma)^{n}\right].\label{S_i}
\end{equation}
At this stage, one may realize the close relation between systematic $1/\text{N}_{f}$ expansion and the RPA series. The extra one in the power of $\mathrm{N}_{f}$ effectively counts the number of closed fermion loops, each of which carries a factor of $\mathrm{N}_{f}$. While in the RPA series, one may just count the number of $\delta\Sigma$ in a diagram, and the number of screened interactions would be half of that.

The change in topology in those two sibling schemes results in drastic difference between the number of diagrams in each order. E.g., at the second order in $1/\text{N}_{f}$ expansion, four $n=3$ $\text{G}-\Sigma$ vertices has six screened interactions and four closed fermion loops, while that can only occur at the sixth order in the RPA series. The pattern in the interaction vertices was then used to build all diagrams.

\section{DATA PREPARATION, LEADING ORDERS AND GRAPHICS} \label{sec:leading_orders}
The first step is to solve the self-consistent Hartree equation. The Dyson's equation for the Matsubara Green's function \footnote{With the momentum and internal band indices hidden and internal matrix multiplication is implicitly indicated. We choose the Fourier transform convention as 
\begin{equation*}
    \bm{c}_{\bm{k}}=\frac{1}{\sqrt{N_{1}N_{2}}}\sum_{\bm{R}_{i}}\bm{c}_{\bm{R}_{i}}e^{i\bm{k}\cdot\bm{R}_{i}},
\end{equation*}
i.e. no internal phase factor between two sites in a unit cell.} with only the Hartree self-energy as 
\begin{equation}
    \left[\frac{\mathrm{d}}{\mathrm{d}\tau}+h\right]G(\tau)=\delta(\tau)+\Sigma_{\text{H}}[G(-\epsilon)]G(\tau).\label{Hartree_Dyson}
\end{equation}
Eq. \eqref{Hartree_Dyson} can be efficiently solved using the \texttt{newton-krylov} method for large number of unit-cells.

In momentum space, the integral equation for the screened interactions is (indices suppressed)
\begin{equation}
\begin{aligned}
    W(\tau,\bm{k})=&-v(\bm{k})\delta(\tau)
    +\\
    &\frac{1}{\sqrt{N_{1}N_{2}}}\int_{\tau^{\prime}}v(\bm{k})P(\tau-\tau^{\prime},\bm{k})W(\tau^{\prime},\bm{k}),
    \label{screened_w}
\end{aligned}
\end{equation}
where the polarization matrix is
\begin{equation}
    P_{b^{\prime}c}(\tau-\tau^{\prime},\bm{k}^{\prime})=\sum_{\bm{k}^{\prime}}G_{\star b^{\prime}c}(\tau-\tau^{\prime},\bm{k})G_{\star,cb^{\prime}}(\tau^{\prime}-\tau,\bm{k}^{\prime}-\bm{k}).
\end{equation}
In practice, it is useful to separate out the singular part of $W(\tau,\bm{k})=-v(\bm{k})\delta(\tau)+\delta W(\tau, \bm{k})$, where $\delta W(\tau, \bm{k})$ satisfies
\begin{equation}
\begin{aligned}
\delta W(\tau,\bm{k})=&-\frac{1}{\sqrt{N_{1}N_{2}}}v(\bm{k})P(\tau,\bm{k})v(\bm{k})\\
&+\frac{1}{\sqrt{N_{1}N_{2}}}v(\bm{k})\int_{\tau^{\prime}}P(\tau-\tau^{\prime},\bm{k})\delta W(\tau^{\prime},\bm{k}).\label{integral_eq_delta_W}
\end{aligned}
\end{equation}

The saddle point contribution to $\mathrm{lnZ}$ is directed evaluated from the action. The $\mathcal{O}(\mathrm{N}_{f}^{0})$ (next order to the saddle point) contribution to $\mathrm{lnZ}$ is evaluated from the functional determinant
\begin{equation}
    \mathrm{lnZ}^{(1)}=-\frac{1}{2}\mathrm{Tr}\,\mathrm{ln}(\mathds{I}-K_{G}),\label{functional_det}
\end{equation}
where $K_{G}(1,2;3,4)=\sumint_{3^{\prime}}v(3^{\prime},3)G_{\star}(3^{\prime},2)G_{\star}(1,3^{\prime})\delta_{3,4}$. RHS of Eq. (\ref{functional_det}) can be graphically represented as
\begin{widetext}
\begin{equation*}
  \begin{tikzpicture}[baseline=(current bounding box.center)]
    \begin{feynman}
      \vertex (d) at (15, 1);  
      \vertex (e) at (15, 0);  

      \diagram*[scale=1]{
        (d) -- [fermion, line width=2, half left] (e),
        (e) -- [fermion, line width=2, half left] (d),
        (d) -- [photon, color=blue] (e)
      };
      \vertex [below=0.3em of e] {\(1\)};
      \vertex [above=0.3em of d] {\(2\)};
    \end{feynman}
  \end{tikzpicture}
  \quad +\;\frac{1}{2}\quad
\begin{tikzpicture}[baseline=(current bounding box.center)]
    \begin{feynman}
      \vertex (d) at (15, 1);  
      \vertex (e) at (15, 0);  
      \vertex (r1) at (15, -1);  
      \vertex (r2) at (15, -2);  

      \diagram*[scale=1]{
        (d) -- [fermion, line width=2, half left] (r2),
        (r2) -- [fermion, line width=2, half left] (d),
        (d) -- [photon, color=blue] (e),
        (e) -- [fermion, line width=2, half right] (r1),
        (r1) -- [fermion, line width=2, half right] (e),
        (r1) -- [photon, color=blue] (r2),
        };

      \vertex [above right=0.3em of e] {\(1\)};
      \vertex [above=0.3em of d] {\(2\)};
      \vertex [below left=0.3em of r1] {\(3\)};
      \vertex [below=0.3em of r2] {\(4\)};
      \end{feynman}
  \end{tikzpicture}
  \quad +\;\frac{1}{3}\quad
\begin{tikzpicture}[baseline=(current bounding box.center)]
    \begin{feynman}
      \vertex (d) at (15, 1);  
      \vertex (e) at (15, 0);  
      \vertex (r1) at (15, -1);  
      \vertex (r2) at (15, -2);  
      \vertex (r3) at (15, -3);  
      \vertex (r4) at (15, -4);  
      \diagram*[scale=1]{
        (d) -- [fermion, line width=2, half left] (r4),
        (r4) -- [fermion, line width=2, half left] (d),
        (d) -- [photon, color=blue] (e),
        (e) -- [fermion, line width=2, half right] (r1),
        (r1) -- [fermion, line width=2, half right] (e),
        (r1) -- [photon, color=blue] (r2),
        (r2) -- [fermion, line width=2, half right] (r3),
        (r3) -- [fermion, line width=2, half right] (r2),
        (r3) -- [photon, color=blue] (r4),
        };

      \vertex [below=0.3em of e] {\(1\)};
      \vertex [above right=0.3em of d] {\(2\)};
      \vertex [below left=0.25em of r1] {\(3\)};
      \vertex [below=0.3em of r2] {\(4\)};
      \vertex [below left=0.3em of r3] {\(5\)};
      \vertex [below=0.25em of r4] {\(6\)};      
      \end{feynman}
  \end{tikzpicture}
  \quad +\;...,
\end{equation*}
\end{widetext}
A trick to evaluate the functional determinant is to introduce an auxiliary variable $x$ such that 
\begin{equation}
    v\sum_{n=0}^{\infty}\frac{1}{n}(P\circ v)^{n}=\int_{0}^{1}v\sum_{n=0}^{\infty}x^{n}(P\circ v)^{n}\;\mathrm{d}x.
\end{equation}
And thus a self-consistent integral equation for $\delta W_{x}(\tau,\boldsymbol{k})$ is formed akin to Eq. \eqref{integral_eq_delta_W}, with $x$ multiplied on the RHS. 

From low to intermediate orders, it is better to adopt the approach that all diagrams are pre-generated. This procedure was benchmarked by checking the weight factor of individual diagrams for Green's functions cancelled with its multiplicity via topological filtering \cite{VF2_1,VF2_2,networkx}, while for free-energy diagrams, symmetry factors do exist and we found that it matched with textbook results for lowest order diagrams. Also, integrals with filtered and unfiltered graphs were compared to be matched. Large graphs matches were employed using the VF2 algorithms \cite{VF2_1, VF2_2,networkx}.

Each edge of the diagram is assigned with a momentum and an imaginary time. We picked up a number of free momentum based on the rank of the momentum conservation equations (similar approach to that used in \cite{Chen_Haule}) and a number of free time based on the fact that time sums to zero for each closed loop, where extra attention has to be paid for the contact/non-contact interaction switch, as in Eq. \eqref{screened_w}.

For sufficient higher orders, the factorial decrease of the average sign becomes a central issue of the simulation \cite{sign_problem_polynomial_complexity}by sampling diagrams individually. Previous approaches employing the determinant trick \cite{RPADet,CoS} with bold-line series may apply to $1/\mathrm{N}_{f}$ expansions due to the additional requirement for partitioning the number of closed fermion loops. We will discuss our combinatorial optimizations later, but focus on preparing building blocks in this section. Non-local vertices in each diagram can be integrated to retrieve the problem of more spread of vertices and use the two-point $\mathcal{L}$ functions as building blocks, defined as: 
\begin{equation}
\begin{aligned}
\mathcal{L}_{abc}(\tau_{1}, \tau_{2}, \bm{k}, \bm{k}^{\prime})=&\sum_{d}\int_{\tau^{\prime}}G_{*,ad}(\tau_{1}-\tau^{\prime}, \bm{k})\\
& G_{*,dc}(\tau^{\prime}+\tau_{2},\bm{k}^{\prime}) W_{db}(\tau^{\prime},\bm{k}-\bm{k}^{\prime}).
\end{aligned}
\end{equation}
Its storage is a practical issue for memory usage, where we used the discrete-Lehmann representation \cite{discrete_lehmann_rep, libdlr}
\begin{equation}
W_{ab}(\tau,\bm{k})=\sum_{l=1}^{r(\beta)}\hat{W}_{ab}(\omega_{l,ab},\bm{k})\frac{e^{-\omega_{l,ab}\tau}}{1+e^{-\beta\omega_{l,ab}}}.
\end{equation}
Since the analytical form of Green's function in momentum space is known, the complexity for single function call only scales as $\mathcal{O}(r)$. We give more details on discrete Lehmann representations and computations of the $\mathcal{L}$ function in Appendix. \ref{app:DLR}. Since bold-line diagrams are not organized in the particle-particle or particle-hole channels as in \cite{RPADet}, the cover by $\mathcal{L}$ is not full, and thus there would be remaining fermion and screened interaction lines left.

\section{EXAMPLES} \label{sec:examples}
\subsection{2D SU(2) Hubbard Model on Square Lattice}
We firstly benchmarked the correctness of the simulation with exact diagonalization for grand-canonical potential density in $3\times3$ square lattice with the $\mathrm{SU}(2)$ Hubbard model, shown in the Fig. 1. 

There is a further large system benchmark compared with the CDet result \cite{Rossi_DDMC}. Several other double-blind parameters in Fig. 1, 
, demonstrating its ability to make realistic predictions.

\begin{singlecolfigure}
\begin{center}
\includegraphics[width=1.0\textwidth]{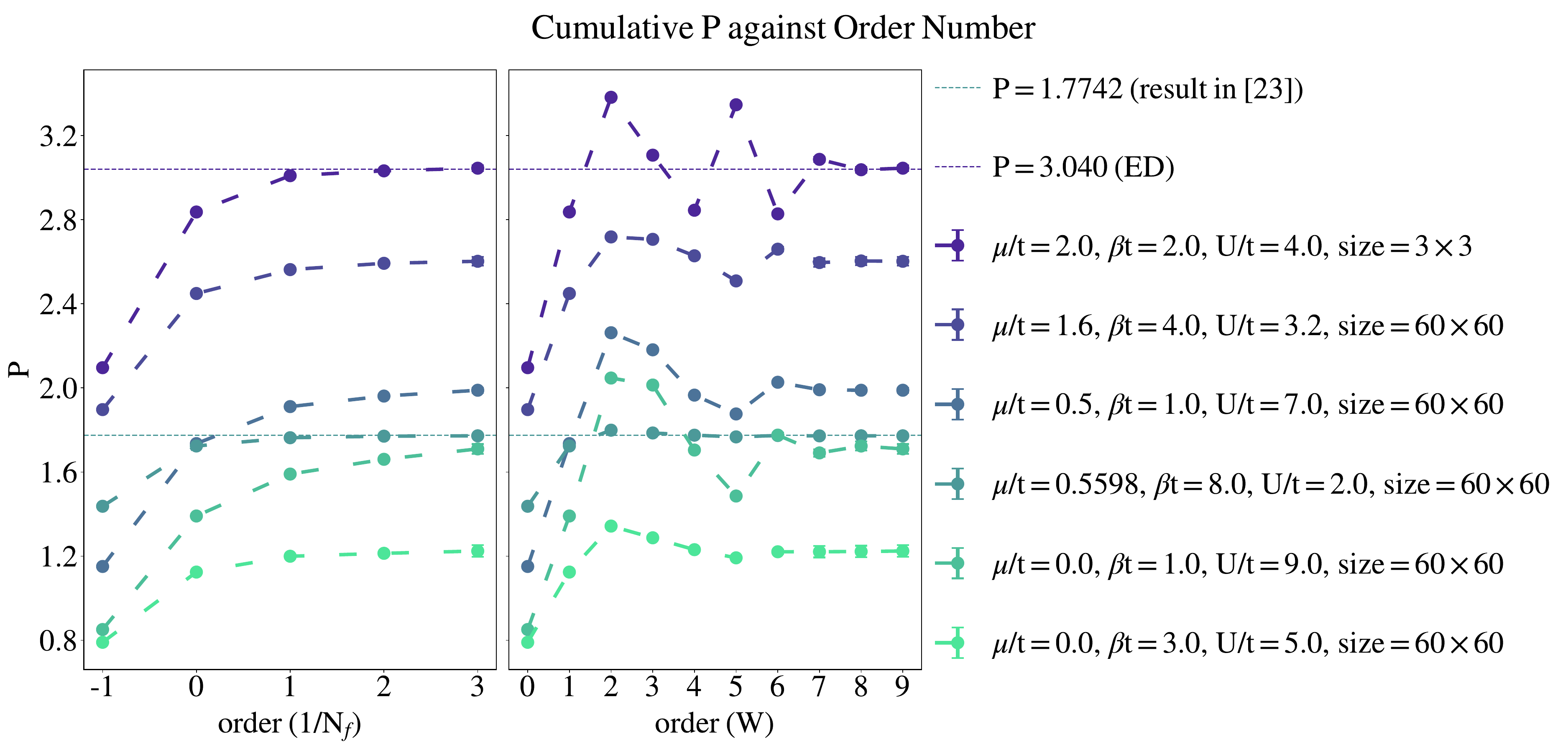}
\twocolcaption{FIG. 1. Cumulative pressure (partial summed pressure cumulating order-by-order contributions) against $1/\mathrm{N}_{f}$ orders and $\mathrm{W}$ orders with $\mathrm{N}_{f}$ cut being $3$. For small systems, pressure is not well defined. We use the definition  $\mathrm{P}=\mathrm{lnZ}/(\beta N_{1}N_{2})$ for plots. Two benchmarks were done with exact diagonalization (ED) for $3\times3$ systems and CDet for $60\times60$ systems drawn from \cite{Rossi_DDMC}. CDet gives \cite{Rossi_DDMC} 1.7742 while $1/\mathrm{N}_{f}$ gives $1.7723(9)$. ED was performed using the \texttt{QuSpin} package \cite{Quspin_1, Quspin_2}. We also simulated a rather high temperature parameter value $\mathrm{U}/\mathrm{t}=5$, $\mu/\mathrm{t}=10$, $\beta\mathrm{t}=0.5$ for $3\times3$ system, which gives $\mathrm{lnZ}=7.704(4)$ compared with exact value $\mathrm{lnZ}=7.7128$. Parameters are shown on the rightmost legend. The LHS plot shows series in powers of $1/\mathrm{N}_{f}$ while RHS plot shows series in powers of $\mathrm{W}$. Order $9$ is the maximal diagrams order with powers of $1/\mathrm{N}_{f}$ less than or equal to $3$, and thus the final result gives the full third-order (with respect to free energy) $1/\mathrm{N}_{f}$ expansion. $-1$ on LHS and $0$ on RHS label the saddle point. Each point in the LHS plot sums several orders in $\mathrm{W}$ series with $\mathrm{N}_{f}$; $\mathcal{O}(1/\mathrm{N}_f)$ spans orders $2$ to $3$ in $\mathrm{W}$, $\mathcal{O}(1/\mathrm{N}^2_f)$ spans orders $3$ to $6$ in $\mathrm{W}$, and $\mathcal{O}(1/\mathrm{N}^3_f)$ spans orders $4$ to $9$ in W.}
\label{Fig:small_large_combined}
\end{center}
\end{singlecolfigure}

We also provide benchmarks at low filling with results in \cite{2DHubbard_Model_Benchmarks}. We consistently found significantly improvements of results using a symmetric Borel-Padé approximation, reflecting the power-law nature of the singularity in the complex $1/\mathrm{N}_{f}$ plane in this regime.

\begin{singlecoltable}
\centering
\renewcommand{\arraystretch}{1.5} 
\setlength{\tabcolsep}{10pt} 
\begin{tabular}{ccccccccccc}
\hline
\hline
\multicolumn{4}{>{\cellcolor[gray]{1.0}}c}{} & \multicolumn{3}{c}{DCA} & 
\multicolumn{2}{c}{DiagMC-$\mathrm{G}\Gamma$} &
\multicolumn{2}{c}{Borel-Padé $\mathcal{O}(1/\mathrm{N}^{2}_{f})$} \\
\cline{5-7}
\cline{10-11}
$\mathrm{t}^{\prime}$ & $n$ & $\mathrm{U}$ & $\beta$ & $\#\;\text{sites}$ & $E$ & $\delta E$ & $E$ & $\delta E$ & $E$ & $\delta E$\\
\hline
0 & 0.3 & 12 & 4 & 20 & -0.821 & 0.013 & -- &--& -0.849 & 0.017 \\

0 & 0.3 & 8 & 4 & 32 & -0.824 & 0.004 &--&--& -0.864 & 0.010 \\

-0.2 & 0.6 & 6 & 2 & 50 & -0.874 & 0.010 &--&--& -0.886 & 0.014 \\

0 & 0.3 & 4 & 4 & 34 & -0.8549 & 0.0002 &-0.8574&0.0007& -0.8605 & 0.007 \\

0 & 0.3 & 6 & 4 & 34 & -0.836 & 0.002 &-0.841&0.002& -0.845 & 0.004 \\
\hline
\hline
\end{tabular}
\twocolcaption{TABLE I. Benchmarks on total energy computations between Borel-Padé $\mathcal{O}(1/\mathrm{N}_{f}^{2})$, (\textit{i.e.} resum the saddle-pt, RPA, $\mathcal{O}(1/\mathrm{N}_{f})$, $\mathcal{O}(1/\mathrm{N}_{f}^{2})$ using the Borel-Padé re-summation with the symmetric rational polynomials.), dynamical cluster approximations (DCA), and G$\Gamma$ diagrammatic Monte Carlo. For DCA, we also give the number of sites of the cluster. $n$ represents the filling fraction and $\mathrm{t}^{\prime}$ is the second-nearest neighbor hopping amplitude.}
\label{tab:connected_vertical}
\end{singlecoltable}

\subsection{2D SU(2) Hubbard model on Honeycomb Lattice}
On honeycomb lattice, we used the state-of-art symmetry broken perturbation series introduced in \cite{resum_techniques, symmetry_broken_perturbation_series} combined with fast principal minor algorithms for cross-benchmark. Since it is for the first time this method is applied, we also give detailed technical steps.

Consider generic bipartite lattices with sub-sites A and B. To increase the radius of convergence, we perform chemical potential plus staggered magnetization shifts, which effectively takes both semi-metal and anti-ferromagnetism behaviors into account. The one-body Hamiltonian becomes (honeycomb lattice with nearest-neighbor hopping)
\begin{equation}
\sum_{\sigma=\uparrow,\downarrow}\bm{c}^{\dagger}_{\sigma}(\bm{k})
\begin{pmatrix}
\Delta_{\sigma}-(\mu-\alpha_{\sigma}) & \epsilon(\bm{k}) \\
\epsilon^{\star}(\bm{k}) & -\Delta_{\sigma}-(\mu-\alpha_{\sigma})
\end{pmatrix}
\bm{c}_{\sigma}(\bm{k}),
\end{equation}
where the hopping energy $\epsilon(\bm{k})=-(1+e^{i\bm{k}\cdot\bm{b}_{1}}+e^{i\bm{k}\cdot\bm{b}_{2}})$ and $\Delta_{\sigma}$, $\alpha_{\sigma}$ are shift parameters. To compensate the shifts, Green's functions with shifted parameters at equal space/time in the determinant (discussed below) are further shifted by 
\begin{equation}
\begin{aligned}
    &g_{\sigma}^{AA}(0^{-},\bm{R}_{i},\bm{R}_{i})=g_{\sigma}^{AA}(0^{-},\bm{R}_{i},\bm{R}_{i})+(\alpha_{\bar{\sigma}}+\Delta_{\bar{\sigma}})/\mathrm{U}\\
    &g_{\sigma}^{BB}(0^{-},\bm{R}_{i},\bm{R}_{i})=g_{\sigma}^{BB}(0^{-},\bm{R}_{i},\bm{R}_{i})+(\alpha_{\bar{\sigma}}-\Delta_{\bar{\sigma}})/\mathrm{U},
\end{aligned}
\end{equation}
where $\bar{\sigma}$ is the opposite spin to $\sigma$. In our numerical examples, we did not choose the shift parameters exactly corresponding to the mean field values to test the robustness of the shifts. Contributions to the partition function $Z$ at order $m$ is (powers of $\mathrm{U}$ suppressed)
\begin{equation}
Z^{(m)}=\frac{1}{m!}\sumint_{X_{1}, ...., X_{m}\atop a_{1},...,a_{m}}\prod_{\sigma}\mathrm{det}
\left[\!\left[
\begin{aligned}
& X_1, \ldots, X_m \\
& X_1, \ldots, X_m
\end{aligned}
\right]\!\right]_{\sigma}^{\{a_{i's}\}},
\end{equation}
where 
\begin{equation}
\begin{aligned}
&\left[\!\left[
\begin{aligned}
X_1, \ldots, X_m \\
X_1, \ldots, X_m
\end{aligned}
\right]\!\right]_{\sigma}^{\{a_{i's}\}}=\\
&
\left|
\begin{array}{ccc}
g_{\sigma}^{a_{1}a_{1}}(X_1, X_1) & \cdots & g_{\sigma}^{a_{1}a_{m}}(X_1, X_m) \\
\vdots & \ddots & \vdots \\
g_{\sigma}^{a_{m}a_{1}}(X_m, X_1) & \cdots & g_{\sigma}^{a_{m}a_{m}}(X_m, X_m)
\end{array}
\right|.
\end{aligned}
\end{equation}
One can readily realize that the summation over all internal sub-lattice sites is just the sum of all the principal minors of the $2m\times2m$ axillary matrix with column and row indices formed by $(X_{1}, A)$, $(X_{1}, B)$, \dots, $(X_{m}, A)$, $(X_{m}, B)$. A simple modification of the fast principal minor algorithm would compute the sub-lattice sum in $\mathcal{O}(2^m)$ operations \cite{fast_principle_minor_DDMC}. Recursive removal of disconnected diagrams follows the standard procedure in \cite{Rossi_DDMC}.

One can directly compute, for example, the spin-up Green's function as well using the following recipe (external points are denoted by $(X, a)$ and $(X', a')$. $a=0,1(A,B)$). First, remove the XOR columns and rows for the spin-up matrix, \textit{i.e.} $M[\bar{a}]=\bm{0}$, $M[:,\;\bar{a'}]=\bm{0}$. (No sum on external sub-lattice indices.). Then, swap the $a$ and $a'$ rows if $a\neq a^{\prime}$. (Principal minors select the same row and columns indices.). In the following, add a $2\times2$ identity block to the upper left of the spin-down matrix. Finally, use exactly the same principal minor summation function to proceed.

We also made comparisons with another way of using fast principal minor algorithms. In the new approach, we leave sub-lattice indices randomly sampled by adding extra variables to the integrand. And the fast principal minors algorithm is only applied to the $m\times m$ matrix spanned by space-time vertices only. We found mostly the new approach is faster since the additional sign problem accumulated by sub-lattice indices are not severe. In practice, we compared results of two approaches for cross-benchmarks.

We found the performance of shifted bare-$\mathrm{U}$ series strongly depending on the shift parameters. We performed benchmarks of this method via exact diagonalization (ED) at $2\times 2$ lattice and at $8\times 8$ lattice with determinant quantum Monte Carlo (DQMC), shown in Appendix. \ref{app:more_bare_U_benchmarks}. We started with rather high temperature at $\mathrm{U}/\mathrm{t}=2$, $\beta\mathrm{t}=2$ for $2\times 2$ lattice and proceeded to low temperature (Appendix. \ref{app:more_bare_U_benchmarks}). Weak coupling results with direct convergence show remarkable accuracy, aligned with general expectations of CDet. 
\begin{singlecolfigure}
\begin{center}
\includegraphics[width=1.0\textwidth]{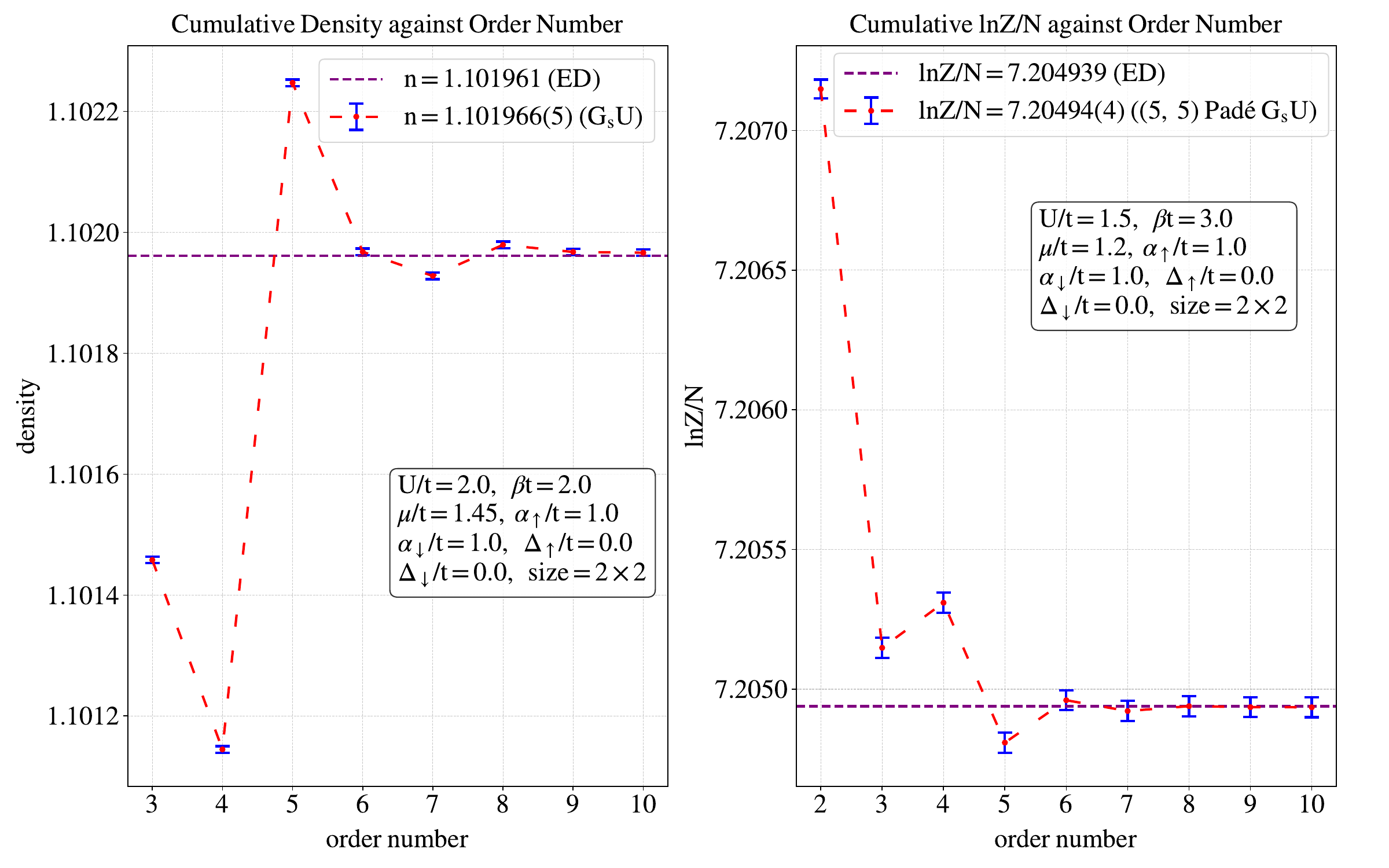}
\twocolcaption{FIG. 2. Cumulative order-by-order density (left), grand potential density (right) on $2\times2$ honeycomb lattice, with all the dynamical and shift parameters given in the text box in the plot. Exact values computed via exact diagonalization for small systems using \texttt{QuSpin} \cite{Quspin_1, Quspin_2} are shown at the top right corner. $\mathrm{G}_{s}\mathrm{U}$ in the legend box means naïve sum of order-by-order corrections.}
\label{Fig:G_0_U_2_2}
\end{center}
\end{singlecolfigure}
In Table. II, 
we provide benchmarks between the $1/\mathrm{N}_{f}$ expansion and determinant quantum Monte Carlo (DQMC). The DQMC simulations were performed using the \texttt{SmoQyDQMC} package in \texttt{Julia} \cite{SmoQyDQMC.jl}. The $\mathcal{O}(1/\mathrm{N}_{f})$ order gives robust correction. Despite only being a toy model, $2\times 2$ systems also exhibit many strongly correlated physics. As can be seen from the top six lines, accuracy when crossing the semi-metal to antiferromagnetic Mott transition is not affected significantly. The same also applies when lowering the temperature. Just as the opposite, we usually found large systems are much easier to be simulated than small clusters due to the smoothness of momentum distribution.

In addition, we also give independent predictions of the grand-potential per site for large system. Pressure measurements to verify the theoretical results are within current cold-atom platforms, though the temperature is challenging to be reached experimentally.
\begin{singlecoltable}
\centering
\renewcommand{\arraystretch}{1.5} 
\setlength{\tabcolsep}{4pt} 
\begin{tabular}{ccccccccccc}
\hline \hline
$\mathrm{U}/\mathrm{t}$ & $\mu/\mathrm{t}$ & $\beta\mathrm{t}$ & Quantity & Size & DQMC/ED & Saddle Pt & RPA & $\mathcal{O}(1/\mathrm{N}_f)$ & $\mathcal{O}(1/\mathrm{N}_f^2)$ & Sum \\
\hline
5 & 1 & 5 & $\mathrm{lnZ}/N$ & $2\times 2$ & 8.498 & 4.9046 & 2.3251 & 1.025(1) & -- & 8.255(1)\\
5 & 2.5 & 5 & $\mathrm{lnZ}/N$ & $2\times 2$ & 15.700 & 9.1325 & 4.6361 & 1.405(3) & -- & 15.174(3)\\
5 & 3.5 & 5 & $\mathrm{lnZ}/N$ & $2\times 2$ & 20.749 & 12.9695 & 6.2621 & 1.481(3) & -- & 20.713(3)\\
6 & 1.5 & 5 & $\mathrm{lnZ}/N$ & $2\times 2$ & 10.402 & 5.5225 & 3.0109 & 1.400(2) & -- & 9.933(2)\\
8 & 4.6 & 8 & $\mathrm{lnZ}/N$ & $2\times 2$ & 40.301 & 20.1859 & 13.2258 & 4.56(5) & -- & 37.97(5)\\
8 & 5 & 12 & $\mathrm{lnZ}/N$ & $2\times 2$ & 65.212 & 33.4798 & 21.8963 & 6.98(5) & -- & 62.36(5)\\
3 & 1 & 2 & density & $4\times 4$ & $0.918$ & 0.6823 & 0.2003 & $0.02966(4)$ & $-0.000025(26)$ & $0.912(1)$\\
4 & -1.42 & 6 & density & $8\times 8$ & $0.355$ & 0.2165 & 0.1103 & $0.02439(6)$ & $0.0026(9)$ & $0.353(2)$ \\
5 & -2 & 5 & density & $8\times 8$ & $0.221$ & 0.1220 & $0.0820$ & $0.02538(7)$ & $0.0042(4)$ & $0.234(1)$ \\
4 & -1 & 6 & $\mathrm{lnZ}/N$ & $50\times 50$ & -- & 1.7069 & 0.8960 & $0.1964(7)$ & $0.0196(33)$ & $2.819(3)$\\
5 & -1 & 8 & $\mathrm{lnZ}/N$ & $50\times 50$ & -- & 1.9594 & 1.2773 & $0.3338(31)$ & $0.051(17)$ & $3.62(2)$\\
7 & 0 & 6 & $\mathrm{lnZ}/N$ & $50\times 50$ & -- &2.6854&2.1085&$0.7836(40)$ & $0.18(13)$& $5.8(1)$\\
\hline \hline
\end{tabular}
\twocolcaption{TABLE II. Benchmarks the $1/\mathrm{N}_{f}$ method with DQMC and exact diagonalization for $\mathrm{SU}(2)$ Hubbard model on honeycomb lattice. For $2\times 2$ systems, we used exact diagonalization while for $4\times4$ and $8\times8$ systems we used DQMC. In DQMC, we have varied trotterization in time ($\Delta \tau=0.1$, $0.05$, $0.03$, $0.025$) and number of bins ($N_{\mathrm{bin}}=50$, $70$) for MCMC blocking to check convergence. We found for density computations where there exhibits fast convergence Borel-Padé re-summation only finely modifies the results.}
\label{Table:honeycomb_benchmarks}
\end{singlecoltable}

\subsection{2D SU($6$) Hubbard Model on Square and Honeycomb Lattices}
We show in Table. III 
comparisons of density computations between the CoS results in \cite{CoS} and the truncated $1/\mathrm{N}_{f}$ results. As the interaction and filling increase, there is a proliferation of error bars in CoS and also in the $1/\mathrm{N}_{f}$ expansion. When both are able to tighten errors at weak coupling, there is a perfect match between the two methods. When it enters the strongly correlated regime, at some filling fractions there are larger differences. We investigated parameter values at $\mathrm{U}/\mathrm{t}=8$, $\mu/\mathrm{t}=1.746251$, $\mathrm{T}/\mathrm{t}=0.15$ and found significant finite-size effects. We conjecture that there is a weak second-order phase transition near this filling. $\mathcal{O}(1/\mathrm{N}_{f})$ expansion is already highly accurate away from the phase transitions and provides decent corrections beyond the RPA results. Importantly, the approximate point of phase transitions can be probed via finite-size effects scaling, where adding the second-order in $1/\mathrm{N}_{f}$ or even higher in perturbation series is required. 

In Fig. 3, 
we also provide equations of states data for $\mathrm{U}/\mathrm{t}=4,6,7,10,13$ at low temperature $\beta\mathrm{t}=7$ on honeycomb lattice. To our knowledge, there have not been systematic studies of this model in the literature. We did not spot drastic changes of series convergence when switching lattice geometry in this regime of parameters. Similar to the square lattice case, compressibility becomes lower when interactions becomes stronger, indicating the Mott tendency.
\begin{singlecolfigure}
\begin{center}
\includegraphics[width=0.7\textwidth]{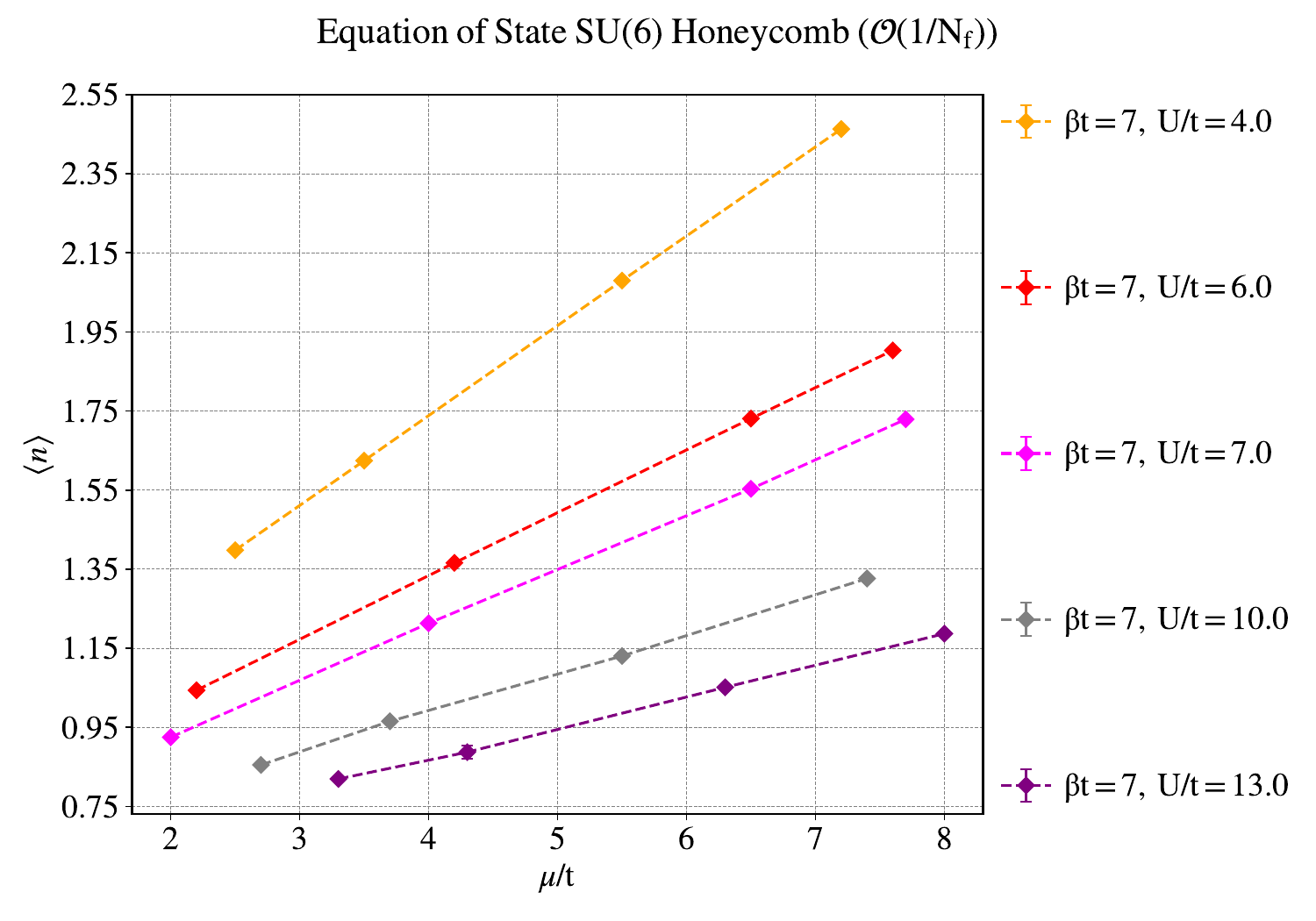}
\twocolcaption{FIG. 3. Equation of state for $\mathrm{SU}(6)$ Hubbard model on $50\times50$ honeycomb lattice for different interacting strength. All the results are obtained by summing the saddle point, RPA and $\mathcal{O}(1/\mathrm{N}_{f})$ orders.}
\label{Fig:EoS_SU_6_honeycomb}
\end{center}
\end{singlecolfigure}

\begin{center}
\begin{singlecoltable}
\centering
\renewcommand{\arraystretch}{1.5} 
\setlength{\tabcolsep}{18pt} 
\begin{tabular}{cccccc}
\hline
\hline
\multicolumn{2}{c}{} & \multicolumn{1}{c}{CoS} & \multicolumn{3}{c}{$1/\mathrm{N}_{f}$} \\
\cline{4-6}
$\mu/\mathrm{t}$ & $\mathrm{U}/\mathrm{t}$ & $n$ & saddle-pt & RPA & Sum $\mathcal{O}(1/\mathrm{N}_{f})$ \\
\hline
-2.300000 & 2.3 & 0.4868(50) & 0.39798 & 0.08224 & 0.487
\\

-1.725000 & 2.3 & 0.6506(5) & 0.53662 & 0.10378 & 0.651
\\

-1.150000 & 2.3 & 0.8178(9) & 0.67783 & 0.12418 & 0.815
\\

0.000000 & 2.3 & 1.1515(12) & 0.96806 & 0.16276 & 1.148
\\

0.575000 & 2.3 & 1.3215(14) & 1.11728 & 0.18126 & 1.318
\\

1.725000 & 2.3 & 1.666(3) & 1.42457 & 0.21729 & 1.665
\\

-1.679207 & 4.0 & 0.5340(25) & 0.38851 & 0.12463 & 0.533(1)
\\

-0.886256 & 4.0 & 0.7021(26) & 0.52397 & 0.15178 & 0.700(1)
\\

-0.088440 & 4.0 & 0.877(4) & 0.66202 & 0.17616 & 0.865(1)
\\

0.714382 & 4.0 & 1.040(5) & 0.80267 & 0.19854 & 1.030(1)
\\

1.522465 & 4.0 & 1.201(6) & 0.94604 & 0.21938 & 1.195(1)
\\

2.336111 & 4.0 & 1.367(7) & 1.09224 & 0.23899 & 1.360(1)
\\

3.155679 & 4.0 & 1.532(12) & 1.24145 & 0.25759 & 1.530(1)
\\

3.981595 & 4.0 & 1.694(16) & 1.39384 & 0.27547 & 1.704(2)
\\

-0.218517 & 8.0 & 0.635(15) & 0.37908 & 0.18969 & 0.623(2)
\\

1.087260 & 8.0 & 0.808(18) & 0.51151 & 0.22081 & 0.781(3)
\\

1.746251 & 8.0 & 0.900(15) & 0.57872 & 0.23438 & 0.866(4)
\\

2.409349 & 8.0 & 0.995(15) & 0.64660 & 0.24695 & 0.941(3)
\\

3.076639 & 8.0 & 1.045(15) & 0.71515 & 0.25867 & 1.021(3)
\\

3.748223 & 8.0 & 1.120(25) & 0.78440 & 0.26967 & 1.098(4)
\\

5.104735 & 8.0 & 1.245(45) & 0.92502 & 0.28991 & 1.263(4)
\\

6.479902 & 8.0 & 1.399(45) & 1.06863 & 0.30825 & 1.417(6)
\\

7.874923 & 8.0 & 1.555(55) & 1.21540 & 0.32510 & 1.566(7)
\\

9.291230 & 8.0 & 1.695(70) & 1.36554 & 0.34081 & 1.759(8)
\\

1.242173 & 12.0 & 0.720(35) & 0.37508 & 0.23023 & 0.674(5)
\\

3.980462 & 12.0 & 0.962(25) & 0.57287 & 0.27468 & 0.92(1)
\\

4.442917 & 12.0 & 0.985(30) & 0.60643 & 0.28085 & 0.95(1)
\\

5.840947 & 12.0 & 1.085(40) & 0.70813 & 0.29787 & 1.06(1)
\\

8.687005 & 12.0 & 1.25(10) & 0.91637 & 0.32685 & 1.30(1)
\\

10.623692 & 12.0 & 1.38(10) & 1.05898 & 0.34347 & 1.44(2)
\\

14.600865 & 12.0 & 1.64(14) & 1.35416 & 0.37237 & 1.75(2)
\\
\hline
\hline
\end{tabular}
\twocolcaption{TABLE III. Density benchmarks for SU(6) Hubbard model on the square lattice with the Hartree-shifted bare-$\mathrm{U}$ series implemented by the CoS algorithm in \cite{CoS}. The $\mathcal{O}(1/\mathrm{N}_{f})$ represents the results by summing saddle-point, RPA and $\mathcal{O}(1/\mathrm{N}_{f})$ correction for $70\times70$ system. Temperature is set at $\mathrm{T}/\mathrm{t}=0.15$, the lowest attempted in \cite{CoS}. The last column represents summing the saddle point, RPA and $\mathcal{O}(1/\mathrm{N}_{f})$ order. Directly summing the fourth and fifth columns would give the saddle point plus RPA corrections.}
\label{tab:SU_6_density}
\end{singlecoltable}
\end{center}
\section{CONCLUSION AND OUTLOOK}
We have proposed a promising numerical method based on $1/\mathrm{N}_{f}$ expansions that is tested for sufficiently large systems with strong interactions at low temperatures for Hubbard-type models. Obvious extensions include changing lattice geometries and interaction types. Generalizing to non-equilibrium that are suited to transport simulations seems feasible but with new problems on optimizing the additional exponential complexity of Keldysh indices summations. To handle the symmetry breaking case, a (conjectured) better way is to expand, not the original $\text{SU}(\text{N}_{f})$ symmetry but artificially broaden the symmetry to $\text{SU}(\text{N}_{f})\times\text{SU}(\text{M})$, where certainly the first symmetry may be void \cite{The_Hubbard_Model}. Importantly, to reach high orders for re-summation, we can expand W rather than $1/\mathrm{M}$. Following methodologies in \cite{The_Hubbard_Model}, we can separate the Hubbard interaction as
\begin{equation}
\frac{1}{N_f} \sum_i\left(\frac{1}{2} V n_i^2-\frac{1}{2} J S_i^2+\frac{1}{2} K \Psi_i^{\dagger} \Psi_i\right),
\end{equation}
where the density-density, magnetic and superconducting channels are separated. $n_i$, $S_i$ and $\Psi_i$ are the corresponding order parameters \cite{The_Hubbard_Model}. Different parametric substitutions from fermion bilinear to two-point $G$ fields can give rise to symmetric breakings at leading orders, which is just the Hartree-Fock mean field theory \cite{The_Hubbard_Model}.

Topologies of diagrams are identical in two approaches. Additional learnable parameters represent mixture between different channels. Since in this paper, we only consider the case of flavour-averages Green's functions, it is impossible to directly compute order parameters, while it is feasible in the mixed channel.

To propose new types of symmetry-broken series beyond the spin-rotation symmetry, semi-bold line series would be the most natural choice out of the dilemma. Old problems such as whether loop current orders can exist in the Hubbard model would require more exotic shifted-expansion point. 

If shifted by symmetry breaking terms is numerically expensive for bold-line series, ideas in quantum Monte Carlo to add a pinning order field \cite{pinning_the_order} may also be applied to DiagMC. We left systematic studies of all of those in the future.
\section{ACKNOWLEDGMENTS}
B. S. is grateful for R. Rossi and J. Kaye for discussions their work. For all the RPA data, we have carefully verified the trotter errors are far smaller than the error bars. For future use of RPA data by other readers, there may be a small difference in the last two digits. Data at high orders were re-checked by RG-MCMC, while all the data in the table and figures are done by VEGAS integration without RG-MCMC unless the two are not within each error bars. A little bias may be generated by purely VEGAS integration for normalization constants but this would be less significant than the systematic errors of the truncated series. B. S. is supported by the Imperial College President's Scholarship. Part of the numerical simulations were performed on the Imperial College HPC cluster. The \texttt{C++} VEGAS map interface was adapted from \cite{VEGAS_Map_github_CIGAR}.

\newpage
\clearpage
\section*{APPENDIX}
\begin{appendices}
\section{More Benchmarks for 2D SU(2) Hubbard Model on the Honeycomb Lattice with Symmetry-Broken Bare U Series} \label{app:more_bare_U_benchmarks}
We provide in the appendix more data of shifted bare-$\mathrm{U}$ series benchmarked with DQMC simulations. In real space, there are well-known problems with wider spread of vertices. To guarantee the validity of numerical results, we compared results with sub-lattices randomly sampled and with sub-lattices summed. We also compared results by varying number of uniform grid points in imaginary time. To reduce memory usage and for better numerical stability, Chebyshev decomposition for shifted bare Green's functions were implemented and we varied the number of Chebyshev polynomials for convergence.
\begin{singlecolfigure}
\begin{center}
\includegraphics[width=1.0\textwidth]{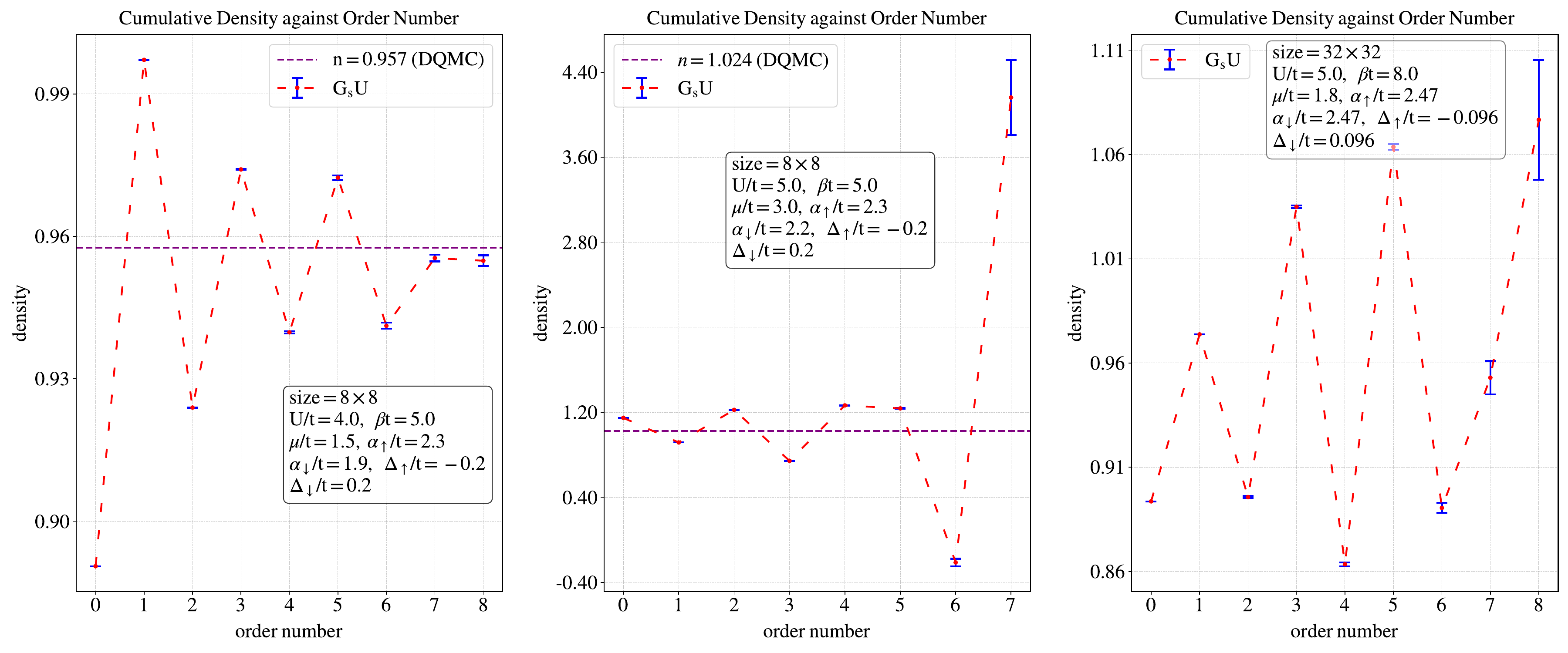}
\twocolcaption{FIG. 4. Cumulative order-by-order density for $8\times8$ and $32\times32$ honeycomb lattice using shifted bare-$\mathrm{U}$ series. Exact value computed DQMC is shown in the legend. Since MC errors at the large expansion orders are large, for LHS plot, $(3,5)$ Padé gives $0.9547(3)$. For middle plot, $(2,5)$ Padé gives $1.04(1)$. For RHS plot for large systems, $(4,4)$ Padé gives $0.964(2)$, $(3,5)$ and $(5,3)$ Padé give $0.961(2)$.}
\label{Fig:DQMC_benchmarks_shifted_bare_U}
\end{center}
\end{singlecolfigure}
As can be seen from the right plot in Fig. 5,
series on the right plot diverges while the Padé approximation (discussed later) correctly predict the filling after re-summation. However, we must point out that at many parameters we have tested, different re-summation methods differ significantly. In particular, singularities on the positive real axis under the Borel-Padé approximation that directly leads to failure of re-summation is absent in the Padé approximation. We find the integral approximant method introduced in \cite{resum_techniques} works poorly in cases with large divergence. The power of re-summation persists to large systems ($50\times 50$). 

\section{More on Discrete Lehmann Representations} \label{app:DLR}
\begin{singlecolfigure}
\begin{center}
\includegraphics[width=0.9\textwidth]{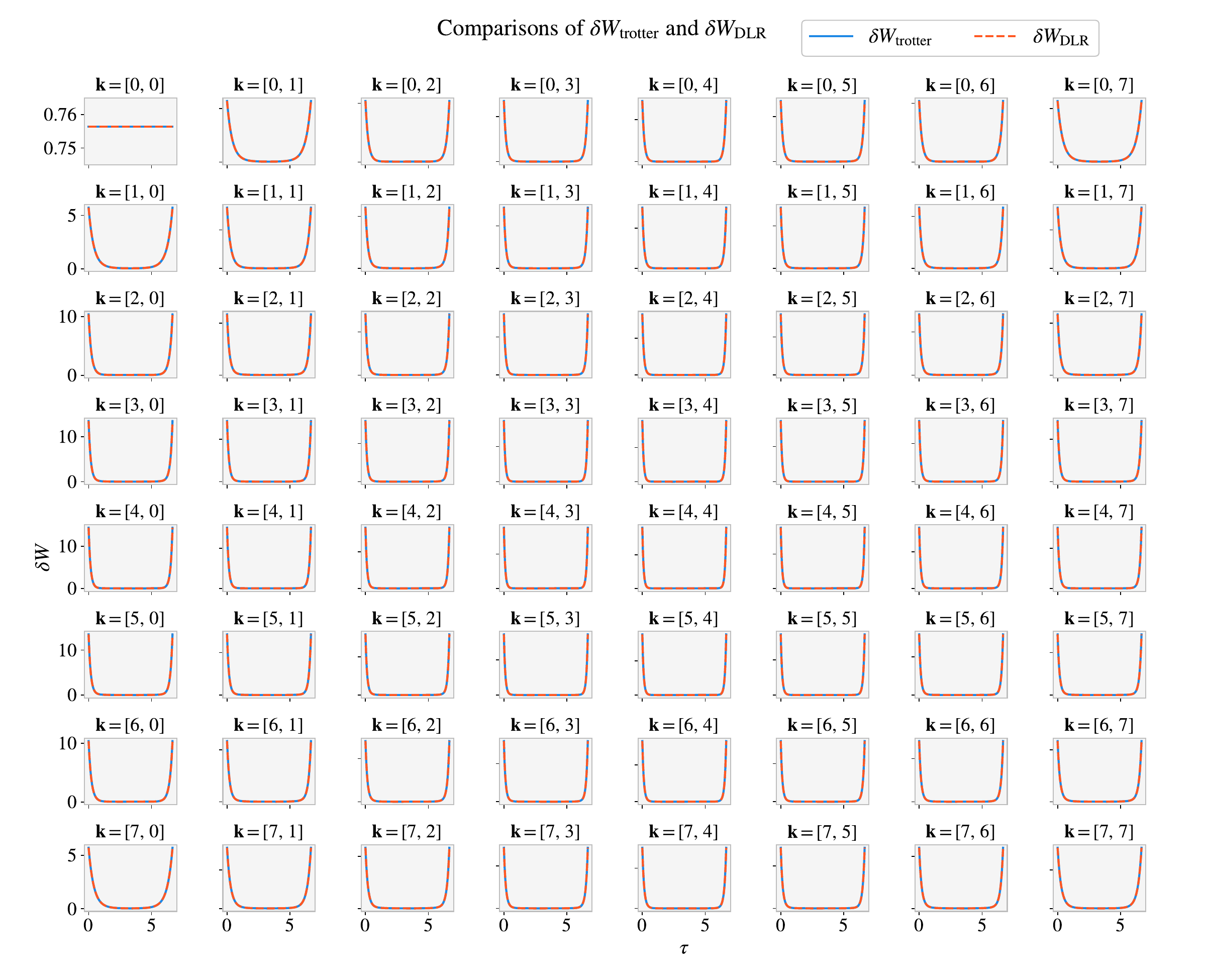}
\twocolcaption{FIG. 5. Comparisons between the data for $\delta W(\tau,\boldsymbol{k})$ and its discrete Lehmann representations for 2D $8\times 8$ SU(6) Hubbard model on the square lattice at $\mathrm{U}/\mathrm{t}=8$, $\mathrm{T}/\mathrm{t}=0.15$ and $\mu/\mathrm{t}=1.746251$. We choose $10000$ points to discretize the uniform mesh in imaginary time and directly solve the integral equation for $\delta W$. After preparing the FFT data, we then perform the DLR decomposition with cut-off $\Lambda=9\beta$ and $\epsilon=10^{-13}$.}
\label{Fig:DLR_delta_W}
\end{center}
\end{singlecolfigure}
With the DLR data, we then proceed to compute the $\mathcal{L}$ function (the contact part is simple to be dealt with and filtered out), 
\begin{equation}
\begin{aligned}
    \mathcal{L}_{abc}(\tau_{1},\tau_2,\boldsymbol{k}_{0},\boldsymbol{k}_{1})=&\sum_d\int_{\tau} G_{ad}(\tau_1-\tau, \boldsymbol{k}_{0})\\
    & G_{dc}(\tau+\tau_{2},\boldsymbol{k}_{1})\delta W_{db}(\tau,\boldsymbol{k}_{0}-\boldsymbol{k}_{1}).
\end{aligned}
\end{equation}
If $\tau_{1}+\tau_{2}<\beta$, it gives
\begin{widetext}
\begin{equation}
\begin{aligned}
V_{a\alpha}V_{\alpha d}^{\dagger} V^{\prime}_{d\gamma} V^{\prime\dagger}_{\gamma c}&\left\{
\frac{e^{(\beta-\tau_{1})\epsilon_{\alpha}+(\beta-\tau_{2})\epsilon^{\prime}_{\gamma}}}{(1+e^{\beta\epsilon_{\alpha}})(1+e^{\beta\epsilon_{\gamma}^{\prime}})}\left(\sum_{l\in\mathcal{D}}p_{l,db}\tau_{1}+\sum_{l\in\bar{\mathcal{D}}}\frac{p_{l,db}}{\epsilon_{\alpha}-\epsilon^{\prime}_{\gamma}-\omega_{l,db}}\left(e^{\tau_{1}(\epsilon_{\alpha}-\epsilon^{\prime}_{\gamma}-\omega_{l,db})}-1\right) \right) \right.\\
&\left. -\frac{e^{-\tau_{1}\epsilon_{\alpha}+(\beta-\tau_{2})\epsilon_{\gamma}^{\prime}}}{(1+e^{\beta\epsilon_{\alpha}})(1+e^{\beta\epsilon_{\gamma}^{\prime}})}\left(\sum_{l\in\mathcal{D}}p_{l,db}(\beta-\tau_{1}-\tau_{2})+\sum_{l\in\bar{\mathcal{D}}}\frac{p_{l,db}}{\epsilon_{\alpha}-\epsilon_{\gamma}^{\prime}-\omega_{l,db}}\left(e^{(\beta-\tau_{2})(\epsilon_{\alpha}-\epsilon^{\prime}_{\gamma}-\omega_{l,db})}-e^{\tau_{1}(\epsilon_{\alpha}-\epsilon^{\prime}_{\gamma}-\omega_{l,db})}\right) \right) \right. \\
&\left. +\frac{e^{-\tau_{1}\epsilon_{\alpha}+(2\beta-\tau_{2})\epsilon_{\gamma}^{\prime}}}{(1+e^{\beta\epsilon_{\alpha}})(1+e^{\beta\epsilon_{\gamma}^{\prime}})}\left(\sum_{l\in\mathcal{D}}p_{l,db}\tau_{2}+\sum_{l\in\bar{\mathcal{D}}}\frac{p_{l,db}}{\epsilon_{\alpha}-\epsilon_{\gamma}^{\prime}-\omega_{l,db}}\left(e^{\beta(\epsilon_{\alpha}-\epsilon^{\prime}_{\gamma}-\omega_{l,db})}-e^{(\beta-\tau_{2})(\epsilon_{\alpha}-\epsilon^{\prime}_{\gamma}-\omega_{l,db})}\right) \right) \right\} \\
\end{aligned}
\end{equation}
If $\tau_{1}+\tau_{2}>\beta$, it gives
\begin{equation}
\begin{aligned}
V_{a\alpha}V_{\alpha d}^{\dagger} V^{\prime}_{d\gamma} V^{\prime\dagger}_{\gamma c}&\left\{
\frac{e^{(\beta-\tau_{1})\epsilon_{\alpha}+(\beta-\tau_{2})\epsilon^{\prime}_{\gamma}}}{(1+e^{\beta\epsilon_{\alpha}})(1+e^{\beta\epsilon_{\gamma}^{\prime}})}\left(\sum_{l\in\mathcal{D}}p_{l,db}(\beta-\tau_{2})+\sum_{l\in\bar{\mathcal{D}}}\frac{p_{l,db}}{\epsilon_{\alpha}-\epsilon^{\prime}_{\gamma}-\omega_{l,db}}\left(e^{(\beta-\tau_{2})(\epsilon_{\alpha}-\epsilon^{\prime}_{\gamma}-\omega_{l,db})}-1\right) \right) \right.\\
&\left. -\frac{e^{(\beta-\tau_{1})\epsilon_{\alpha}+(2\beta-\tau_{2})\epsilon_{\gamma}^{\prime}}}{(1+e^{\beta\epsilon_{\alpha}})(1+e^{\beta\epsilon_{\gamma}^{\prime}})}\left(\sum_{l\in\mathcal{D}}p_{l,db}(\tau_{1}+\tau_{2}-\beta)+\sum_{l\in\bar{\mathcal{D}}}\frac{p_{l,db}}{\epsilon_{\alpha}-\epsilon_{\gamma}^{\prime}-\omega_{l,db}}\left(e^{\tau_{1}(\epsilon_{\alpha}-\epsilon^{\prime}_{\gamma}-\omega_{l,db})}-e^{(\beta-\tau_{2})(\epsilon_{\alpha}-\epsilon^{\prime}_{\gamma}-\omega_{l,db})}\right) \right) \right. \\
&\left. +\frac{e^{-\tau_{1}\epsilon_{\alpha}+(2\beta-\tau_{2})\epsilon_{\gamma}^{\prime}}}{(1+e^{\beta\epsilon_{\alpha}})(1+e^{\beta\epsilon_{\gamma}^{\prime}})}\left(\sum_{l\in\mathcal{D}}p_{l,db}(\beta-\tau_{1})+\sum_{l\in\bar{\mathcal{D}}}\frac{p_{l,db}}{\epsilon_{\alpha}-\epsilon_{\gamma}^{\prime}-\omega_{l,db}}\left(e^{\beta(\epsilon_{\alpha}-\epsilon^{\prime}_{\gamma}-\omega_{l,db})}-e^{\tau_{1}(\epsilon_{\alpha}-\epsilon^{\prime}_{\gamma}-\omega_{l,db})}\right) \right) \right\},
\end{aligned}
\end{equation}
\end{widetext}
where $V$ and $V^{\dagger}$ are unitary matrices diagonalizing the momentum space free Green's function, $p$ and $\omega$ follow from conventions in the main text,
where 
\begin{equation}
    \delta W_{db}(\tau,\boldsymbol{k})=\sum_{l}p_{l}(\boldsymbol{k})\;e^{-\omega_{l,db}\tau}
\end{equation}
To simplify notations, $V$ and $\epsilon$ symbols with $\prime$ are for $\boldsymbol{k}_1$; $V$ and $\epsilon$ symbols without $\prime$ are for $\boldsymbol{k}_0$. $\mathcal{D}$ and $\bar{\mathcal{D}}$ denotes resonant and non-resonant sets of $l$ when $\epsilon-\epsilon^\prime-\omega=0\;\mathrm{or}\neq0$. To handle the summation of large exponential, it would be better to re-organize terms in terms of product of fermi functions
\begin{equation}
\begin{aligned}
& f_{F}(\tau)=-e^{-\tau\epsilon}/(1+e^{\beta\epsilon})\;\mathrm{if}\;\tau\leq0,\\
& f_{F}(\tau)=e^{(\beta-\tau)\epsilon}/(1+e^{\beta \epsilon})\;\mathrm{if}\;\tau>0.
\end{aligned}
\end{equation}
We need to reintroduce the fermi factor back in $\delta W$ to ensure we can fully factorize into the product of fermi functions:  
\begin{equation}
    \delta W_{db}(\tau,\boldsymbol{k})=\sum_{l}p_{l}(\boldsymbol{k})\;\frac{e^{-\omega_{l,db}\tau}}{1+e^{-\beta\omega_{l,db}}}
\end{equation}
\section{List of Parameters for MCMC} \label{app:MCMC_para}
We give the full list of parameters and their typical values used for controlling the accuracy of the simulation. We choose a uniform mesh in the imaginary time $[\epsilon,\beta-\epsilon]$, where a small $\tau$-grid shift $\epsilon$ is to avoid discontinuities at the boundaries. We introduced an auxiliary $x$ variable to compute $\mathrm{ln}Z_{\mathrm{RPA}}$ and numerical integration over $x$ are performed using trapezoidal rule with $N_{x}$ points in $[0, 1]$. For the discrete Lehman representation, we reduce the error by tuning tolerance ($\mathrm{DLR}_{\mathrm{tol}}$) and cut-off ($\mathrm{DLR}_{\Lambda}$). We use central differentiation to compute density and energy 
\begin{equation}
\begin{aligned}
& n=\frac{\mathrm{ln}Z(\mu+h_{\mu})-\mathrm{ln}Z(\mu-h_{\mu})}{2N h_{\mu}},\\
& E=-\frac{\mathrm{ln}Z(\beta+h_{\beta})-\mathrm{ln}Z(\beta-h_{\beta})}{2N h_{\beta}}+\\
&\quad\quad\;\mu \frac{\mathrm{ln}Z(\mu+h_{\mu})-\mathrm{ln}Z(\mu-h_{\mu})}{2N\beta h_{\mu}},
\end{aligned}
\end{equation}
where $N$ is the total number of sites. To avoid MC error bar proliferation, we found it is advantageous to directly perform the numerical differentiation within the integrand. For VEGAS map, we also introduce a damping factor $\alpha_{\alpha}$ for the original VEGAS damping $\alpha$ during map iterations to gradually reduce $\alpha$, where we found too high $\alpha$ values could lead to degenerate grids and could not stabilize the map. Parameters values with -- means they vary largely from case to case. 
\begin{singlecoltable}
\centering
\renewcommand{\arraystretch}{1.5} 
\setlength{\tabcolsep}{15pt} 
\begin{tabular}{ccc}
\hline
\hline
\textbf{Param} & \textbf{Symbol} & \textbf{Typical Value} 
\\ \hline
Number of $\tau$ grid points & $N_{\mathrm{UM}}$ & 5000\\

$\tau$ grid shifts & $\epsilon$ & $10^{-14}$\\

Number of $x$ grid points & $N_{x}$ & 30\\
 
$\mathrm{DLR}\;\mathrm{tol}$ & 
$\mathrm{DLR}_{\mathrm{tol}}$ & --\\

$\mathrm{DLR}\;\mathrm{\Lambda}$ & 
$\mathrm{DLR}_{\mathrm{\Lambda}}$ & --\\

Chemical Potential Shift & $h_\mu$ & 0.01 \\

Inverse Temperature Shift & $h_\beta$ & 0.02 \\ 

MCMC Warm-up Periods & $N_{\mathrm{warm}}$ & $2^{16}$ \\ 

MCMC Updates & $N_{\mathrm{MC}}$ & -- \\ 

VEGAS training per iteration & $N_{\mathrm{VEGAS}}$ & -- \\ 

Number of VEGAS iterations & $N_{\mathrm{iter}}$ & $64$ \\ 

Number of MC Samplings for $\mathcal{N}$ & $N_{\mathcal{N}}$ & -- \\ 

Number of VEGAS Intervals & $N_{I}$ & $100$ \\ 

VEGAS Damping $\alpha$ & $\alpha$ & $0.8$ \\ 

$\alpha$ for VEGAS Damping $\alpha$ & $\alpha_{\alpha}$ & $1.1$ \\ 

Number of MCMC Chains & $N_{\mathrm{chains}}$ & -- \\ 
\hline
\hline
\end{tabular}
\twocolcaption{\qquad\qquad\qquad\qquad\qquad\quad\quad\; TABLE IV. List of parameters to control the simulation accuracy.}
\label{tab:random_data}
\end{singlecoltable}
\clearpage
\newpage
\section{Statistics for RG-MCMC}
\label{app:RG_MCMC}
As a practical example for RG-MCMC, we demonstrate with the energy data for $50\times 50$ square lattice at order $6$ with partition $(3,3,3,3)$ at $\mathrm{U}/\mathrm{t}=8$, $\beta\mathrm{t}=4$, $\mu/\mathrm{t}=-1.77$. We used three chains, with the first one in linear dimension: $50\to20$, the second in $\beta\mathrm{t}$: $4\to3\to2.5\to2\to1.5\to1.0$, and the third in $\mathrm{U}/\mathrm{t}$: $8\to7\to6\to5\to4\to3$. The final estimate is done by a simple MC integration with parameters at the end of each chain. We plot the QQ-plot for all the chains in Fig. 7. 
As a comparison, the leftmost subplot in Fig. 7 
is done by the VEGAS integration.
\begin{singlecolfigure}
\begin{center}
\includegraphics[width=0.9\textwidth]{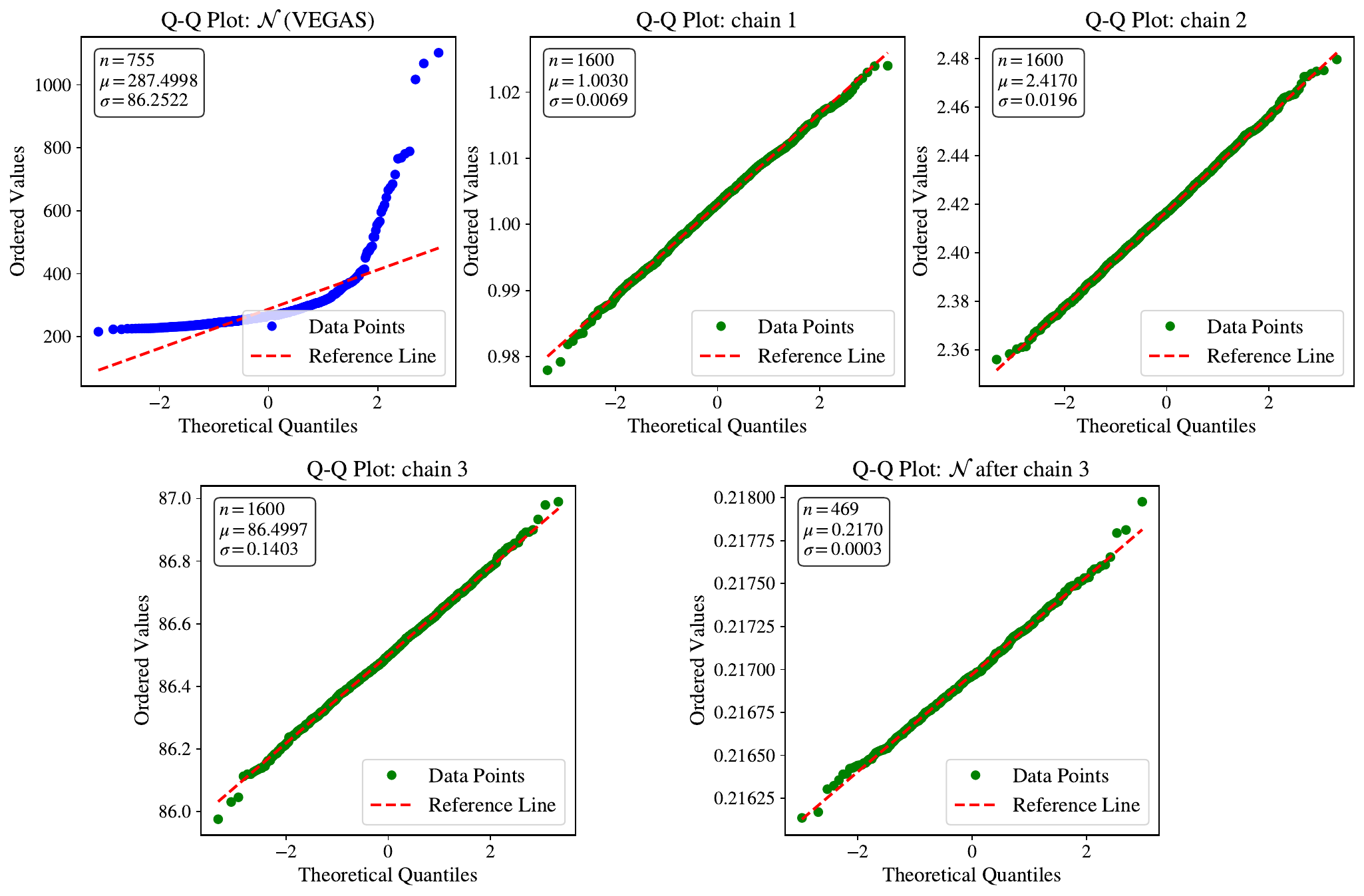}
\twocolcaption{FIG. 6. QQ-plot for all the three chains of the exemplary simulation. Each data point represents results after a single MCMC chain (or VEGAS MC integration) with random seeds completely different. $n$ is the number of data points, $\mu$ and $\sigma$ are mean and std. The rightmost plot applies the VEGAS integration only where there are significant deviations from the Gaussian distribution. The RG-MCMC method significantly improves this, where the total number of function evaluations are roughly the same. Despite the visually large deviations and very low $p$-value, we just find a slight bias in the error bar estimate with the VEGAS integration for the final result.}
\label{Fig:qq_plot}
\end{center}
\end{singlecolfigure}
\end{appendices}
\providecommand{\noopsort}[1]{}\providecommand{\singleletter}[1]{#1}%

\end{document}